# Intrusion Detection Systems for Smart Home IoT Devices: Experimental Comparison Study

**Faisal Alsakran[1], Gueltoum Bendiab[1], Stavros Shiaeles[2], and Nicholas Kolokotronis[3]**

[1] CSCAN, University of Plymouth, PL4 8AA, Plymouth, UK

faisal.alsakran@postgrad.plymouth.ac.uk, bendiab.kelthoum@umc.edu.dz

[2] School of Computing, University of Portsmouth, PO1 2UP, Portsmouth, UK

sshiaeles@ieee.org

[3] Department of Informatics and Telecommunications, University of Peloponnese,

22131 Tripolis, Greece

nkolok@uop.gr

## ABSTRACT

Smart homes are one of the most promising applications of the emerging Internet of Things (IoT) technology. With the growing number of IoT related devices such as smart thermostats, smart fridges, smart speaker, smart light bulbs and smart locks, smart homes promise to make our lives easier and more comfortable. However, the increased deployment of such smart devices brings an increase in potential security risks and home privacy breaches. In order to overcome such risks, Intrusion Detection Systems are presented as pertinent tools that can provide network-level protection for smart devices deployed in home environments. These systems monitor the network activities of the smart home-connected de-vices and focus on alerting suspicious or malicious activity. They also can deal with detected abnormal activities by hindering the impostors in accessing the victim devices. However, the employment of such systems in the context of smart home can be challenging due to the devices hardware limitations, which may restrict their ability to counter the existing and emerging attack vectors. There-fore, this paper proposes an experimental comparison between the widely used open-source NIDSs namely Snort, Suricata and Bro IDS to find the most appropriate one for smart homes in term of detection accuracy and resources consumption including CP and memory utilization. Experimental Results show that Suricata is the best performing NIDS for smart homes.

---

[*]Alternative email bendiab.kelthoum@umc.edu.dz

[†]Member of the Cyber Security Research Group

## 1 INTRODUCTION

Smart home technology also often referred to as home automation allows the entire home to be automated and therefore, the connected smart home devices can be remotely controlled and operated, from any location in the world, through a smartphone app, iPads or other network devices [13]. In recent years, smart home technology is gaining tremendous ground at all levels. Economic reports affirm that connected home market becomes the largest IoT segment at seven billion related smart devices in 2018, which present 26% of the global IoT devices market [14]. According to Gartner [32] this segment is expected to grow to 20.4 billion devices by 2020. Further, the number of householders with smart systems has grown to nearly 150 million smart householders' worldwide in 2019 [14]. The main reasons for the large adoption of such technology are comfort, convenience, safety, and energy and cost savings [13].

However, connecting smart devices such as lights, appliances and locks introduces tremendous cybersecurity risks. All security reports warn that more than 80% of connected smart home devices are vulnerable to a wide range of attacks [11, 25]. A recent research reported by the cybersecurity firm Avast affirms that two out of five smart households are vulnerable to cyberattacks [5]. Exploiting such unsecured devices by hackers can lead to all kinds of potential harm [11, 17], like switching the security system to unlock a door [11], or cracking the smart oven until overheats and burns the house down [11]. In other cases, the smart home network is infected with ransomware that requires the homeowner to pay in order to regain access to the home network [25]. Even a simple smart light bulb can be exploited by hackers to gain wider access to the smart home network and cause potential physical damage [17].

In light of all of this, it is clear that there is a major gap be-tween security requirements and security capabilities of currently available smart home IoT devices. One of the main reasons that make these devices insecure is the hardware limitations [4, 30]. More specifically, restricted resources including low power sources, small amounts of memory and limited processing power, which means minimizing the number of processes, and consequently, the size of the applications. These limitations hinder the execution of complexes security tasks that generate massive computation and communication load [4]. Consequently, security solutions for these devices should maintain a balance between the smart home high-security requirements and supporting infrastructures' hardware limitations. As this new technology has a direct impact on our life's security and privacy, this issue must become a higher priority for security and home automation experts [18]. In this context, there is a need for efficient Intrusion Detection Systems (IDSs), which can provide high protection for smart devices deployed in home environments with a minimum of resources consumption. In fact, IDSs employment in the context of smart home can be challenging due to the device's hardware limitations [18, 25].

In this paper, we aim to address this issue by examining the existing IDSs in order to find the most appropriate one for smart homes in term of accuracy and resources consumption. In this con-text, several pen-source network-based intrusion detection systems (NIDS) are increasingly being used as they offer many benefits and are freely available, such as ACARM-ng, AIDE, Bro IDS, Snort, Suri-cata, OSSEC HIDS, Prelud Hybrid IDS, Samhain, Fail2Ban, Security Onion, etc. These systems are considered as a cost-effective way to improve the security of smart home environments by monitoring the home network and detect internal or external cyber-attacks [26]. However, in this experimental study we will focus on Snort, Suricata and Bro-IDS. Many studies show that these three NIDSs are the most efficient and become the de-facto industry standard for intrusion detection engines [2, 23, 27, 29]. The main contribution of this paper is a performance comparison of those three IDSs based on some significant features including accuracy, CPU and RAM utilisation. The chosen IDSs are deployed inside different Linux containers instead of running them directly on the VM base operating system. Each container has its own resources that are isolated from other containers. By doing this, Snort, Suricata and Bor-IDS will be deployed on the same virtual machine with a minimum of resources. Therefore, conducting experiments will be easier. The experiments evaluate the difference in resource usage between these NIDSs while monitoring live network traffic under various attacks.

The rest of the paper is structured as follows. In section II, we briefly review some previous work that is related to our work. Section III gives an overview of the chosen IDSs Snort, Suricata and Bro. Section IV explains our evaluation experiments and the results, and Section V concludes the paper and outlines the potential future work.

## 2 RELATED WORK

In recent years, researchers have increased their interests in studying the performance of different NIDSs in different environments, from different perspectives. In this context, the performance of the Snort IDS has been extensively investigated in research studies [9, 22, 28, 29]. For instance, in [28] authors conducted experimental evaluation and comparison of the performance of Snort NIDS when running under the two popular platforms Linux and Windows. The evaluation is done for both normal and malicious traffic, and the metrics used were the throughput and the packet loss. Those experiments showed that Snort is performing better on Linux than on Windows. In another work [22], authors examined the performances of snort 2.8.0 NIDS by considering CPU and memory usage, system bus, network interface card, hard disc, logging technique, and the pattern matching algorithm.

This study shows that hardware resources have a clear impact on the overall snort IDS performance. While authors in [9] studied the limitations of snort IDS by conducting several experiments on a real network environment. The metrics used to analysis the Snort performance are the number of packets received, the number of packets analysed, the number of packets filtered, and the number of packets dropped. The study showed that the Snort IDS failed to process high-speed network traffic and the packet drop rate was higher.

Several other studies conducted performance comparison between the two popular open IDS systems Snort and Suricata [1, 3, 8, 34]. In [3], authors investigated the performance of Snort and Suricata on three different platforms: ESXi virtual server, Linux 2.6 and FreeBSD. The experiments were carried out for different packet sizes and speeds and measure the rates of packet drop and alerts. Authors reported that Suricata gave better performance on Linux, while FreeBSD is the ideal platform for Snort especially when the later run on the high-speed network traffic. In 1],[ the performance comparison study of Snort and Suricata IDSs focused on identifying the computer host resource utilisation performance and detection accuracy. The experiments were carried out on two different computer hosts with different CPU memory and network card configurations. This study showed that Snort requires less processing power to perform well compared to Suricata. However, Suricata is more accurate in detecting malicious traffic with high computing resources and its ruleset is more effective. In another recent study [8], authors analysed and compered Snort and Suricata performances on Windows and Linux platforms. Experiment results showed that both IDSs use more resources on the Linux operating system. Authors concluded that CPU usage is highly affected by the operating system on which the IDS is deployed for both solutions. Study in [15] reached the same conclusions as in [8]. Authors reported that Linux-based execution of both IDSs consumes more system resources than its windows-based counterpart. With a similar intention, the study in [29] investigated the performance of Snort and Suricata for accurately detecting the malicious traffic on computer networks. The performance of both IDSs was evaluated on two computers with the same configuration, at 10 Gbps network speed. Authors concluded that Suricata could process a higher speed of network traffic than Snort with lower packet drop rate, but it consumed higher computational resources. In [10], authors focused on packet drops. They found that both Snort and Suricata performed with the same behaviour with larger packets and larger throughputs.

Few studies have considered other IDSs in the comparison such as in [31], where authors studied the performance and the detection accuracy of Snort, Suricata and Bro. The evaluation is done using various types of attacks (DoS attack, DNS attack, FTP attack, Scan port attack and SNMP attack), under different traffic rates and for different sets of active rules. The performance metrics used are the CPU utilization, the number of packets lost, and the number of alerts. In this study, Bro IDS showed better performance than Suricata and snort when evaluated under different attack types for some specific set of rules. Also, authors concluded that the high traffic rate has a significant effect on the CPU usage, the packets lost and the number of alerts for the three IDSs. In a previous work [24], author compared the three above-mentioned IDSs, looking for advantages and disadvantages of each one.

The evaluation was performed at different network speeds. The experimental results showed that Suricata and Bro IDSs can handle 100 Mbps and 1 Gbps network speeds with no packet drops. In a similar context, authors in [33] proposed a new methodology to assess the performance of the intrusion detection systems snort, Ourmon and Samhain in a simulated environment. The simulation experiments were con-ducted on physical and virtual machines to measure the CPU load, memory need, bandwidth constraint and computer memory in-put/output. Authors concluded that Snort imposes more impact on network traffic than Ourmon and Samhain IDSs. In [19] a high-level comparison is done between Snort and Bro. In this study, the authors affirmed that Snort is the best lightweight IDS but it not good for high-speed networks. Whereas Bro is more effective for Gbps networks, but it is more complex to deploy and understand. In more recent work [26] authors provided a high-level analysis and performance evaluation of popular IDSs including Snort, Suricata, Bro IDS, Open WIPS-ng, OSSEC, Security Onion and Fragroute. The survey concluded that most of the existing IDSs have low detection accuracy with minimum hardware and sensor support.

## 3 OVERVIEW OF SNORT, SURICATA AND BRO IDS

To identify threats, Network-based intrusion detection systems (NIDS) collect information about incoming and outgoing traffic from the internet (Figure 1) [1, 12]. These systems utilise a combination of signature-based and anomaly-based detection methods. Signature-based detection involves comparing the collected data packets against signature files that are known to be malicious, while anomaly-based detection method uses behavioural analysis to monitor events against a baseline of "normal" network activity. When malicious activity arises on a network, NIDSs detect the activity and generate alerts to notifying administrators or blocking the source IP address from reaching the network [12].

There are several open-source NIDS/NIPS engines available to automate and simplify the process of intrusion detection, and Snort is one of the best solutions for small and lightly utilized networks [27].

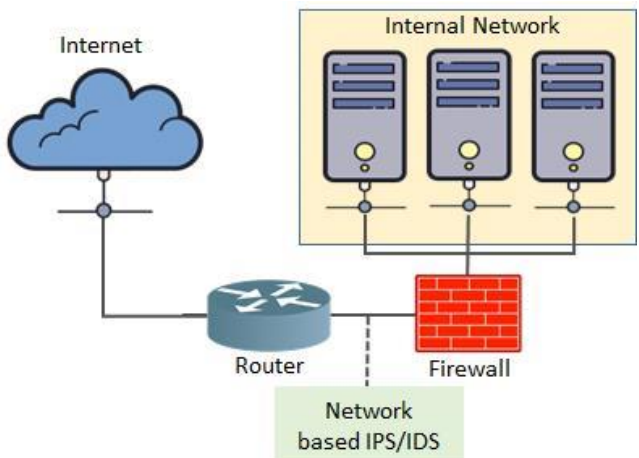

Figure 1: IDS/IPS in a network architecture.

It was developed in 1998 by Martin Roesch from Sourcefire [1] and is now owned by Cisco, which acquired Sourcefire in 2013 [6]. Snort is the most widely deployed IDS/IPS worldwide over the last decades [29]. According to The Snort website, this IDS has been downloaded over 5 million times so far and currently has more than 600, 000 registered users [6]. It has single-threaded architecture, which uses the TCP/IP stack to capture and inspect network packets payload [21, 29]. However, their last version Snort 3.0 has added the multiple packet processing threads in order to address the limitation of single-threaded architecture in the previous versions. It uses both signature-based (SIDS) and anomaly-based (AIDS) methods for anomaly detection.

The Suricata IDS is a relatively new NIDS compared to Snort, it was developed in 2010 by the Open Information Security Foundation (OISF)[2] in an attempt to meet the requirements of modern infrastructures [29]. This NIDS introduced multi-threading to help speed up the network traffic analysis and overcome the computational limitations of single-threaded architecture [20, 31]. Like Snort; Suricata is rules-based and offers compatibility with Snort Rules [29], it also provides intrusion prevention (NIPS) and net-work security monitoring [8], and uses both signature-based and anomaly-based methods to detect malicious network traffic [31]. Unlike Snort, Suricata provides offline analysis of PCAP files by using a PCAP recorder. It also provides excellent deep packet inspection and pattern matching which makes it more efficient for threat and attack detection [1]. The industry considers Suricata a strong competitor to Snort and thus they are often compared with each other.

Bro-IDS is an open-source Unix-based NIDS and passive network traffic analysis [35]. It was originally developed in 1994 by Vern Paxson and renamed Zeek in late 2018 [35]. Bro IDS work differently from Snort and Suricata because of its focus on network analysis. It works as NIDS by passively monitors the network traffic and looks for suspicious activity [23]. Also, Bro policy scripts are written in its own Bro scripting language that does not rely on traditional signature detection. Further, Suricata and snort are under GNU GPL licence [29], support IPv6 traffic and their installation and deployment are easy [29]. In contrast, Bro-IDS is under BSD license, does not support IPv6 traffic and their installation can be difficult [29, 31, 34]. In fact, Bro is more difficult and consume more time to deploy and to understand [7]. Further, Snort and Suricata can run on all operating systems (e.g., Linux, Mac OS X, FreeBSD, OpenBSD, UNIX and Windows) and not restricted to a fully vested server hardware platform whereas Bro is confined to UNIX like operating systems, which hinders their portability. Like snort and Suricata, Bro IDS also uses both signature-based intrusion and anomaly-based methods to detect unusual network behaviour [7, 31].

Table 1 shows a high-level comparison between the three IDSs and gives an overview of the different parameters can be assembled. This high-level comparison shows that the three intrusion detection systems have their benefits and there is no system with a clear advantage over the others.

---

[1] Sourcefire: www.sourcefire.com

[2] OISF: https://suricata-ids.org/about/oisf/

F. Alsakran et al.Table 1: Comparison table of Snort, Suricata and Bro IDSs

| Parameters | Snort | Suricata | Bro IDS |
|---|---|---|---|
| Provider | Cisco System | OISF | Vern Paxson |
| Open-source licence | GNU GPL licence | GNU GPL licence | BSD license |
| Operating system | Win/Unix/Mac | Win/Unix/Mac | Unix/FreeBSD |
| Installation/deployment | Easy | Intermediate | Typical |
| Intrusion prevention capabilities | Yes | Yes | No |
| Network Traffic | IPv4/IPv6 | IPv4/IPv6 | IPv4 |
| Intrusion detection technique | SIDS, AIDS | SIDS, AIDS | SIDS, AIDS |
| Configuration GUI | Yes | Yes | No |
| Support to high-speed network | Medium | High | High |

## 4 EXPERIMENTAL METHODOLOGY

As mentioned before, smart homes security becomes a challenging topic, in which the security and home automation experts try to maintain a balance between the smart home high security requirements and supporting infrastructures' hardware limitation. In general, these environments suffer from inherent hardware limitations, which restrict their ability to implement comprehensive security measures and increase their exposure to vulnerability attacks. To select the appropriate security solutions, it is indispensable to examine these hardware limitations and make sure that they will not affect the performance of these solutions in protecting the smart home-related devices. In light of this, we aim in these experiments to examine the well-known intrusion detection systems Snort 3.0, Suricata 3.0.1 and ID Bro 2.5 to find the most suitable one for smart homes in term of resources consumption and the detection accuracy. More concretely, we examined the real-time performances of these IDSs while monitoring live network traffic from the smart home. Performance information from the CPU and RAM will be recorded, analysed and compared.

### 4.1 Experimental setup

The experiments were performed on a virtual machine running Ubuntu 16.0.x OS, with 8 GB of RAM, 40 GG of HDD and Intel Xeon CPU E5-2650 v2 running at 2.6 GHz. In the simulation scenarios, we first take a snapshot of the clean machine before executing any malicious sample. Then, after executing the malicious sample and recorded all information related to resources consumption and VM state, the VM is reverted to its original form. In order to emulate the smart home environment, we used Docker Enterprise (EE) to run the three IDSs inside Linux containers than running them directly on the VM base operating system. Docker showed great superiority when compared to normal VMs or hypervisors in terms of disk and memory management, start-up and compilation speed, and overall processing performance [16]. In these experiments, each IDS was separately installed on identical custom Docker containers with default performance parameters (Figure 2).

The performance evaluation of each IDS is done for 20 PCAP samples of malicious traffic generated by different types of attacks. The PCAP files were collected from (malware-traffic-analysis.net).

The same malicious pcap files were used to monitor the resources used by the three IDSs while doing analysis of traffic and generating alerts. For the performance evaluation of the three IDSs, the information recorded during the execution of the malicious pcap samples include CPU and RAM use. TCPreplay is used to replay the malicious pcap files to the NIDSs (Figure 2). Table 2 shows the PCAP samples of malicious traffic used in the experiments.

Table 2: PCAP samples of malicious traffic

| #Id | Type of the malware in the PCAP file | Size of the PCAP file |
|---|---|---|
| #1 | Malspam traffic | 1.0 3MB |
| #2 | Necurs Botnet Malspam | 448 KB |
| #3 | Payment Slip Malspam | 3.0 MB |
| #4 | MedusaHTTP malware | 669 kB |
| #5 | Adwind Malspam | 1.7 kB |
| #6 | KainXN EK | 1.93 MB |
| #7 | Cyber Ransomware | 584 KB |
| #8 | Locky-malspam-traffic | 285 KB |
| #9 | Facebook Themed Malspam | 1.4 MB |
| #10 | BOLETO Malspam infection traffic | 2.4 MB |
| #11 | Pizzacrypt | 254.4 KB |
| #12 | BIZCN Gate Actor Nuclear | 0.9 8 MB |
| #13 | Fiesta Ek | 1.52 MB |
| #14 | Nuclear EK | 2 MB |
| #15 | Fake-Netflix-login-page-traffic | 768 KB |
| #16 | URSNIF Infection with DRIDEX | 2.5 MB |
| #17 | Dridex Spam trrafic | 999 KB |
| #18 | Brazil malware spam | 12.7 MB |
| #19 | Info stealer that uses FTP to exfiltrate data | 1.4 MB |
| #20 | Hookads-Rig-EK-sends-Dreambot | 595 KB |

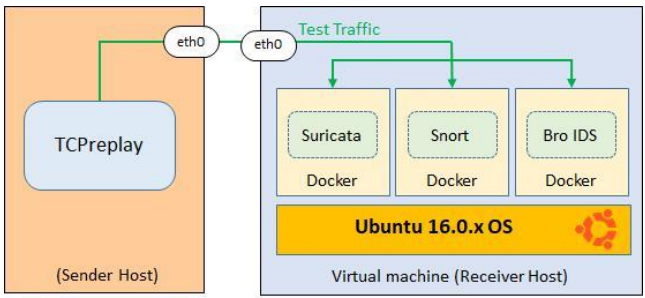

Figure 2: Overview of the Testbed.

### 4.2 Experiments Results

#### 4.2.1 RAM utilisation.

Figure 3 compares the results for the RAM utilisation rate for each malware sample, for the three IDSs Snort, Suricata and Bro IDS. From the obtained results, we can also have the same conclusions for the CPU usage; the Snort IDS gives the highest RAM utilisation rates for most of the PCAP samples. The rates are in the range of 60% and approximately 80%. While the highest rates were recorded for samples #4, #13 and #19. Suricata recorded relatively lower rates than Snort ranged from 20% to approximately 40%. While Bro IDS was the Best IDS in term of RAM usage by recording the lowest rates for most of the tests (From 20% to approximately 34%). Like in the CPU tests, it is also observed that the type of malware traffic has a significant effect on the RAM usage for the three IDSs.

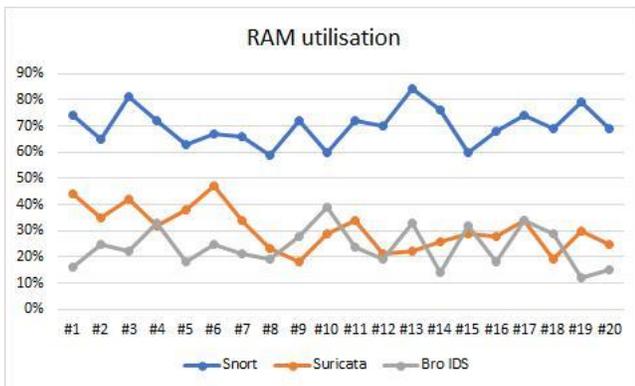

Figure 3: RAM utilisation results.

#### 4.2.2 CPU utilisation

Figure 3 compares the results for the CPU utilisation rate for each malware sample, for the three IDSs Snort, Suricata and Bro IDS. From the obtained results, it is observed that the Snort IDS recorded the highest CPU utilisation rates for most of the PCAP samples, between 60% and 70%. While Suricata and Bro recorded relatively lower rates for the same malware attack tests. The CPU utilisation for Both IDSs is ranging from 20% to 40%, however, Suricata gives the lower rates for most of the tests compared to Bro-IDS and Snort. It is also observed that the type of malware traffic has a significant effect on the CPU usage, each IDS gives different CPU usage rates for the same test attack as they act quite differently for each attack.

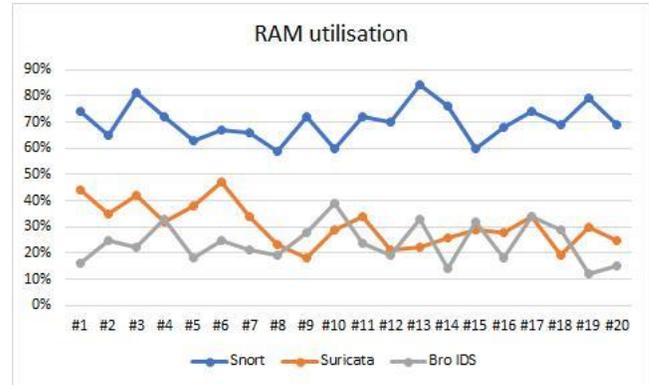

Figure 4: CPU utilisation results.

In summary, it can be concluded from this quantitative comparison of the three ISDs, in term of resource usage (CPU and RAM), that Snort utilisation of CPU and memory was higher than that of Suricata and Bro. The reasons for that are the usage of Dockers and the support of multiple packet processing threads architecture in this version of Snort (Snort3), which require more computational resources compared to previous versions of Snort. Suricata used an average of 30.5% of memory, which exceeded Snort's memory utilisation by approximately 10%, whereas the two IDSs achieved approximately the same results in term of CPU usage, with an average usage of 36% for Suricata and 32% for Bro. The obtained results from these experiments demonstrate that Suricata and Bro perform better than Snort 3 in case of hardware limitations, therefore, they are more suitable for smart homes.

## 5 CONCLUSION

In this paper, we compared the performance of the open-source IDS systems Snort, Suricata and Bro to find the most suitable one for smart homes in term of resources consumption including CPU and memory usage. This study using Dockers, showed that each system had its strengths and weaknesses and the experimental results demonstrated that Suricata and Bro utilised less resources than Snort 3, which make them more appropriate to smart homes than Snort 3. In the future, we expect to improve this work by conducting more experiments on the three IDSs in term of detection accuracy as well as resources using larger pcap _les. Finally, we are intended to use real smart home environment to perform the experiments.

## ACKNOWLEDGMENTS

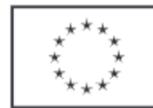

This project has received funding from the European Union's Horizon 2020 research and innovation pro-gramme under grant agreement no. 786698. The work reflects only the authors' view and the Agency is not responsible for any use that may be made of the information it contains.